\documentclass[reprint,amsmath,amssymb,aps,pra,showpacs]{revtex4-1}

\pdfoutput=1

\usepackage{graphicx}
\usepackage{dcolumn}
\usepackage{bm}
\usepackage{hyperref}
\usepackage[caption=false]{subfig}
\usepackage{amsmath}

\hypersetup{
	colorlinks = true,
	urlcolor   = blue,
	linkcolor  = blue,
	citecolor  = blue
}

\renewcommand*{\phi}{\varphi}
\renewcommand*{\epsilon}{\varepsilon}
\renewcommand*{\le}{\leqslant}

\DeclareMathOperator{\Tr}{Tr}
\DeclareMathOperator{\rank}{rank}

\DeclareMathOperator{\prob}{prob}

\newcommand*{\ket}[1]{| #1 \rangle}

\newcommand*{\dyad}[1]{| #1 \rangle \langle #1 |}
\newcommand*{\argmax}{\mathop{\mathrm{argmax}}}

\begin{document}

\title{Experimental adaptive quantum state tomography based on rank-preserving transformations}

\author{A.\,D.\,Moiseevskiy}
	\email{amoiseevskiy@quantum.msu.ru}
\author{G.\,I.\,Struchalin}
\author{S.\,S.\,Straupe}
\author{S.\,P.\,Kulik}
\affiliation{Quantum Technology Centre, and Faculty of Physics, M.\,V.\,Lomonosov Moscow State University, 119991, Moscow, Russia}

\date{\today}

\begin{abstract}
Quantum tomography is a process of quantum state reconstruction using data from multiple measurements. An essential goal for a quantum tomography algorithm is to find measurements that will maximize the useful information about an unknown quantum state obtained through measurements. One of the recently proposed methods of quantum tomography is the algorithm based on rank-preserving transformations. The main idea is to transform a basic measurement set in a way to provide a situation that is equivalent to measuring the maximally mixed state. As long as tomography of a fully mixed state has the fastest convergence comparing to other states, this method is expected to be highly accurate. We present numerical and experimental comparisons of rank-preserving tomography with another adaptive method, which includes measurements in the estimator eigenbasis and with random-basis tomography. We also study ways to improve the efficiency of the rank-preserving transformations method using transformation unitary freedom and measurement set complementation.
\end{abstract}


\maketitle

\section{Introduction\label{sec:Introduction}}
Quantum state tomography is now ubiquitous in the practice of any quantum experimentalist. The goal of this procedure is to provide an estimate $\hat\rho$ of a density matrix for an unknown quantum state $\rho$, given the measurement data~\cite{Paris_Book2004}. The measurements have to be performed in several bases and need to form a tomographically complete set to allow for unambiguous determination of all the matrix elements of the density matrix. Since the measurement outcomes are probabilistic, one needs to perform them on many identically prepared copies of the quantum system and register many outcomes to reduce the statistical uncertainty of the estimation. The ultimate goal of quantum state tomography protocol design is to maximize the precision of estimation and minimize the required experimental resources.

A traditional and still most common practice is to use tomographic protocols with a fixed set of measurements which is independent of the estimated state, we will call such protocols \emph{static}. A static family also includes protocols, where measurements are chosen at random. Static protocols are a good choice in experimental scenarios where only a limited set of measurement operators may be realized, or where changing the measurement basis is a challenging procedure. 

An important figure of merit for protocol optimization is the total number of measurement outcomes $N$ one has to register to reach the desired level of estimation accuracy. It is relevant in the experimental scenario where the low event generation rate is the limiting factor. For static protocols, one may try to optimize the choice of measurement operators to improve the accuracy for a given $N$ providing statistically efficient protocols \cite{Kulik_PRL10}. The ultimate limits on statistical efficiency are known~\cite{Massar_PRA00,Bagan_PRL2006} and correspond to the infidelity $1-F(\hat\rho,\rho)$ between the estimate and the true state scaling as $\alpha/N$, where the prefactor~$\alpha$ depends on the dimensionality and rank of the system. These limits are only achievable with \textit{adaptive} tomographic protocols, where the choice of measurements depends on the information obtained from the previous outcomes. 

There are numerous variants of adaptive strategies for quantum tomography (see~\cite{Straupe_JETP2016} for a review), and most of them provide quadratic improvement in the infidelity scaling approaching the theoretical optimum~\cite{Massar_PRA00,Houlsby_PRA12,Steinberg_PRL13,Guo_NPJQI16,Murao_PRA12,Takeuchi_PRL12,Hen_NJP15,Ferrie_PRL14,Granade_NJP2017}. However the search for simple and computationally inexpensive adaptive procedures for state estimation continues and new protocols are proposed continuously. One of such proposals was recently made in~\cite{Bogdanov_SPIE2019} with a quite extravagant analogy to Lorenz transformations and claims of statistical ``super-efficiency". Here we experimentally investigate this protocol and compare it to the known adaptive strategies in terms of the infidelity scaling with~$N$.

\section{Theory\label{sec::Theory}}
Quantum measurements are characterized by a POVM---positive operator-valued measure~$\mathcal M$. For \emph{projective} measurements in some $D$-dimensional basis, a POVM is defined as a set of~$D$ rank-1 projectors onto vectors~$\ket{\phi_{\gamma}}$ from this basis: $\mathcal M = \{\dyad{\phi_{\gamma}}\}_{\gamma = 1}^{D}$. By definition, POVM elements $M_{\gamma} \in \mathcal M$ are Hermitian non-negative operators that sum to unity:
\begin{equation}
\sum_{\gamma} M_{\gamma} = 1_D, \label{eq:POVMNormalization}
\end{equation}
where~$1_D$ is a $D$-dimensional identity matrix. Note, that a single-basis POVM is not \emph{tomographically complete}, i.\,e., it is insufficient to determine an unknown state uniquely. Therefore, several bases should be concatenated into one complete POVM.

The probability~$p_{\gamma}$ of obtaining an outcome~$\gamma$ with the system being in a state~$\rho$ can be found via the Born rule:
\begin{equation}
p_{\gamma} = \Tr(M_{\gamma} \rho). \label{eq:BornRule}
\end{equation}

\subsection{Unknown state estimation}
One of the methods to estimate an unknown quantum state~$\rho$ is the \emph{maximum likelihood estimation} (MLE)~\cite{Hradil97}. The point of the method is to introduce a function~$\mathfrak L(\mathcal D, \rho)$ that determines the probability of obtaining an outcome sequence~$\mathcal D$ if the system is in a state~$\rho$: $\mathfrak L(\mathcal D, \rho) \equiv \prob(\mathcal D | \rho)$. The state~$\hat \rho = \argmax_\rho \mathfrak L(\mathcal D, \rho)$ that provides the maximum of~$\mathfrak L$ for the measured data~$\mathcal D$ is accepted as an estimator.

For example, in experiments with photon counting, where each measurement~$M_j$ has an exposition time~$t_j$, and assuming Poissonian counting statistics, the likelihood function can be expressed as a product of Poissonian terms:
\begin{equation}
\mathfrak{L}(\mathcal D, \rho) = \prod_{j} \frac{(I p_j(\rho) t_j)^{n_j}}{n_{j}!} \exp{[-I p_j(\rho) t_j]}. \label{eq:PoissonLikelihood}
\end{equation}
Here $I$~is the source intensity, $n_j$ is the number of detected photons for the measurement~$M_j$, and $\mathcal D = \{n_j\}_j$. In further numerical simulations and experimental data analysis we use precisely this definition of the likelihood. The maximum of the likelihood function can be found by an iterative algorithm as described in Ref.~\cite{Bogdanov_JETP2009}.

The difference between density matrices is usually quantified by \emph{fidelity} $F(\rho, \hat \rho) = \Tr^2 \sqrt{\rho^{1/2} \hat \rho \rho^{1/2}}$. Given the fidelity, metrics on the state space can be constructed. One possible choice is the squared \emph{Bures distance}~\cite{Bures69}:
\begin{equation}
d_B^2(\rho, \hat \rho) = 2 - 2 \sqrt{F(\rho, \hat \rho)} \approx 1 - F(\rho, \hat \rho), \label{eq:BuresDistance}
\end{equation}
where the last approximate equality holds for small distances, $d_B^2 \ll 1$.

\subsection{Rank-preserving transformations\label{sec:Transformations}}
In this section, we review the adaptive quantum state estimation protocol proposed in Ref.~\cite{Bogdanov_SPIE2019}. The authors of the original paper rely on an isomorphism between a qubit Bloch vector and an energy-momentum 4-vector of a material point. They use a Lorentz-boost transformation to find measurement operators for a qubit case and then generalize the protocol to higher dimensions. However, we do not resort to Lorentz maps and reformulate the protocol in terms of \emph{rank-preserving transformations}.

Let us start with the fact that for any given \emph{static} protocol, the most precise reconstruction of the true state is achieved when this state is the maximally mixed one~\cite{Bogdanov_SPIE2019}: $\rho_* \equiv 1_D/D$, where $D$ is the Hilbert space dimension. Hence, the main idea of the rank-preserving-transformation protocol is to change the measurements such that the outcome probabilities~$p_\gamma$ are equal to the case of measuring a fully mixed state. To fulfill this condition, we need to transform the measurements from ${M_\gamma}$ to ${M_\gamma^\text{new}}$ in the following way:
\begin{equation}
\Tr(M_\gamma^\text{new} \rho) = \Tr(M_\gamma \rho_*). \label{eq:MainIdea}
\end{equation}

POVM elements~$M_\gamma$ are positive by definition, so their transformation can be written as a completely positive (CP) map of a general form~\cite{Krauss_book_83}:
\begin{equation}
M_\gamma^\text{new} = \sum_i L_i M_\gamma L_i^\dagger, \label{eq:GeneralMTransform}
\end{equation}
where $L_i$ are called operator elements or Krauss operators. The choice of~$L_i$ for a given map is not unique. Therefore, a \emph{Krauss rank} of a map is defined as the minimal possible number of~$L_i$.

Usually, experimenters deal with projective rank-1 measurements, where $M_\gamma = \dyad{\phi_\gamma}$ (our experiment is neither an exclusion). Hence, we consider only this class of measurements throughout the paper. Now we show that if the map~\eqref{eq:GeneralMTransform} always produces rank-1 outputs~$M_\gamma^\text{new}$ for all rank-1 inputs~$M_\gamma$, then its Krauss rank is equal to one, and vice versa. Indeed, if $M_\gamma = \dyad{\phi_\gamma}$, then each term of the sum is a projector onto a vector $L_i \ket{\phi_{\gamma}}$. The output $M_\gamma^\text{new}$ will have unity rank, iff vectors $L_i \ket{\phi_{\gamma}}$ span the same one-dimensional subspace, i.\,e., $\forall\,i, j: L_i \ket{\phi_\gamma} = c_{ij} L_j \ket{\phi_\gamma}$, where $c_{ij}$ are some constants. Provided arbitrary $\ket{\phi_\gamma}$, the last equality holds, iff $L_i = c_{ij} L_j$---all operators differ from each other only by a constant factor. Therefore, they can be replaced by a single operator~$L$, meaning that the map~\eqref{eq:GeneralMTransform} has unity rank:
\begin{equation}
M_\gamma^\text{new} = L M_\gamma L^\dagger. \label{eq:MTransform}
\end{equation}

Since the trace action allows for cyclic permutations of arguments, after substitution of~\eqref{eq:MTransform} into~\eqref{eq:MainIdea} we get:
\begin{multline}
\Tr (M_\gamma^\text{new} \rho) = \Tr([L M_\gamma L^\dagger] \rho) =\\
\Tr(M_\gamma [L^\dagger \rho L])  = \Tr (M_\gamma \rho_*),
\end{multline}
from which we find
\begin{equation}
\rho_* = L^\dagger \rho L = \mathcal L \rho \mathcal L^\dagger, \label{eq:RhoTransform}
\end{equation}
where $\mathcal L = L^\dagger$ is an adjoint operator. The obtained expression imposes additional constraints on~$L$ because the density matrices~$\rho$ and $\rho_*$ have a full rank: $\rank \rho = \rank \rho_* = D$. Indeed, the rank of a product does not exceed the ranks of factors: $\rank AB \le \min(\rank A, \rank B)$ for some square matrices $A$ and $B$. By applying this inequality to Eq.~\eqref{eq:RhoTransform},
\begin{equation}
\rank \rho_* = D \le \min(\rank \rho, \rank L) = \rank L \le D,
\end{equation}
we conclude with the necessity that $\rank L = D$. Therefore, the desired transformation~\eqref{eq:MTransform} or \eqref{eq:RhoTransform} includes a multiplication by a full-rank matrix~$L$, thus, it preserves the rank of the argument, and hence its name.

To find an explicit form of $\mathcal L$ we need a spectral decomposition of the density matrix $\rho$:
\begin{equation}
\rho = U \Lambda U^\dagger, \label{eq:SpectralDecomposition}
\end{equation}
By combining Eqs.~\eqref{eq:RhoTransform} and~\eqref{eq:SpectralDecomposition} we get:
\begin{equation}
\rho_* = 1_D / D = \mathcal L U \Lambda U^\dagger \mathcal L^\dagger.
\end{equation} 
Finally, a general expression for the desired transformation matrix is:
\begin{equation}
\mathcal L = \frac{1}{\sqrt{D}} \frac{1}{\sqrt{\Lambda}} U^\dagger. \label{eq:GeneralTransformation}
\end{equation}

Let us summarize the flow of rank-preserving-transformations protocol:
\begin{enumerate}
	\item\label{itm:Start} Using the recipe~\eqref{eq:GeneralTransformation}, find a transformation~\eqref{eq:RhoTransform} that brings a current maximum likelihood estimator~$\hat{\rho}$ to the fully mixed state~$\rho_*$.
	\item\label{itm:MTransform} Apply the adjoint transformation~\eqref{eq:MTransform} to the basic measurement operators~$M_\gamma$ to obtain the new set~$M_\gamma^\text{new}$.
	\item Measure~$M_\gamma^\text{new}$ and update MLE~$\hat \rho$ according to the detected outcomes.
	\item Repeat the procedure (go to step~\ref{itm:Start}) until all state copies are measured, otherwise stop.
\end{enumerate}
We emphasize that the transformation in step~\ref{itm:MTransform} depends on the current estimator~$\hat \rho$. As tomography proceeds, $\hat \rho$ approaches the true state~$\rho$, therefore, the measurements $M_\gamma^\text{new}$ adapt to the true state.

An interesting analogy between the obtained transformation in the case of qubits and Lorentz transformations of the special relativity theory was noticed in Ref.~\cite{Bogdanov_SPIE2019}. Let us identify the Bloch vector components $s_i$ of the state~$\rho$ together with $\Tr \rho = 1$ with an energy-momentum 4-vector of a corresponding relativistic particle $\{\Tr\rho,s_1,s_2,s_3\}$. Then the transformation~\eqref{eq:RhoTransform} of the state is equivalent to a Lorentz boost from a lab reference frame to the particle rest frame. Accordingly, measurement projectors are transformed as if we looked at them from the reference frame, associated with a relativistic particle traveling in the direction indicated by the Bloch vector representing the estimator~$\hat\rho$.

Note that a fully mixed state~$\rho_*$ is invariant under unitary transformations. So we have the freedom to left-multiply $\mathcal L$ by an arbitrary unitary matrix~$V$:
\begin{gather}
\rho_* = V \rho_* V^\dagger = V \mathcal L \rho \mathcal L^\dagger V^\dagger = \mathcal L' \rho \mathcal L'^\dagger, \nonumber\\
\mathcal L' = V \mathcal L. \label{eq:UnitaryFreedom}
\end{gather}
This unitary freedom is equivalent to a rotation of the initial measurement set as a whole: $M'_\gamma = V^\dagger M_\gamma V$. In our simulations, we tried to analyze the effect of multiplication by random unitary matrices on tomography convergence but observed no significant influence outside the statistical error range.

\subsection{Complementation to the decomposition of unity\label{sec:Complementation}}
In general, the transformations described in Section~\ref{sec:Transformations} are not unitary, and the modified set of measurements $M_\gamma^\text{new} = L M_\gamma L^\dagger$ can violate the POVM normalization requirement:
\begin{equation}
S \equiv \sum_\gamma M_\gamma^\text{new} \neq 1_D. \label{eq:POVMNormalizationViolance}
\end{equation}
This raises a question, how the set $M_\gamma^\text{new}$ can be complemented to recover a decomposition of unity.

We studied two possible approaches. Corresponding protocol modifications will be designated by RankP-B and RankP-M (the letter B stands for ``basis'', M---for ``minimal''). The first one (RankP-B) is to complement each measurement operator~$M^\text{new}_\gamma \equiv \dyad{\phi_{\gamma,1}}$ to a random basis $\{\dyad{\phi_{\gamma,j}}\}_{j=1}^{D}$. This procedure results in an increasingly overcomplete measurement set with growing dimensionality. Another way (RankP-M) is to append a single basis~$\{M_j^\text{cmpl}\}_{j=1}^D$ to the whole measurement set~$\{M^\text{new}_\gamma\}_\gamma$ as follows. Firstly, we normalize the original measurements by the maximal eigenvalue~$\mu_\text{max}$ of the sum~$S$~\eqref{eq:POVMNormalizationViolance}: $M'_\gamma = M_\gamma^\text{new}/\mu_\text{max}$. Then we calculate a spectral decomposition of the difference between the identity matrix and the scaled sum:
\begin{equation}
1_D - \frac{S}{\mu_\text{max}} = \sum_{j=1}^D \lambda_j \dyad{\phi_j}.
\label{eq:ComplementationExplained}
\end{equation}
Finally, additional measurements are expressed as follows:
\begin{equation}
M_j^\text{cmpl} = \lambda_j \dyad{\phi_j}.
\label{eq:Complementation}
\end{equation}
The obtained measurement set again satisfies the POVM normalization requirement: $\sum_\gamma M_\gamma' + \sum_j M_j^\text{cmpl} = 1_D$.

We note that due to the non-unitarity of the transformation to the modified measurements $M_\gamma^\text{new}$ and following~\eqref{eq:POVMNormalizationViolance}, it becomes impossible to estimate the total intensity~$I$ of a photon source only from the measured counts if $M_\gamma^\text{new}$ are not complemented. Therefore, we have to independently measure the intensity~$I$ to be able to estimate an unknown quantum state via the likelihood function~\eqref{eq:PoissonLikelihood}.

\subsection{Normalization and measurement time\label{sec:MeasTimes}}
\begin{figure}
	\centering
	\subfloat[Initial view.]
	{
		\includegraphics[width=0.46\linewidth]{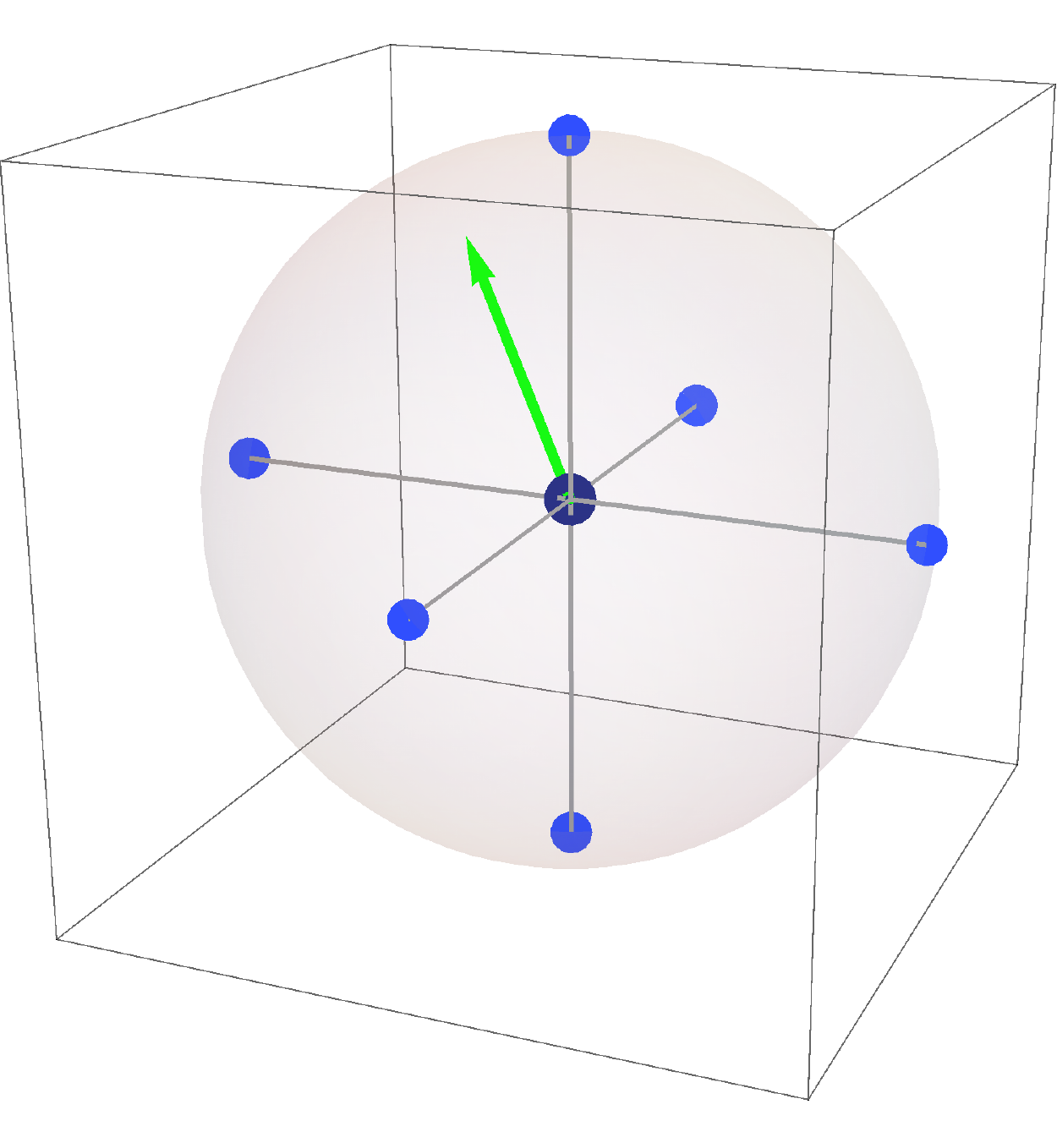}
		\label{fig:NormalizationOk}
	}
	\subfloat[View after a single transformation.]
	{
		\includegraphics[width=0.5\linewidth]{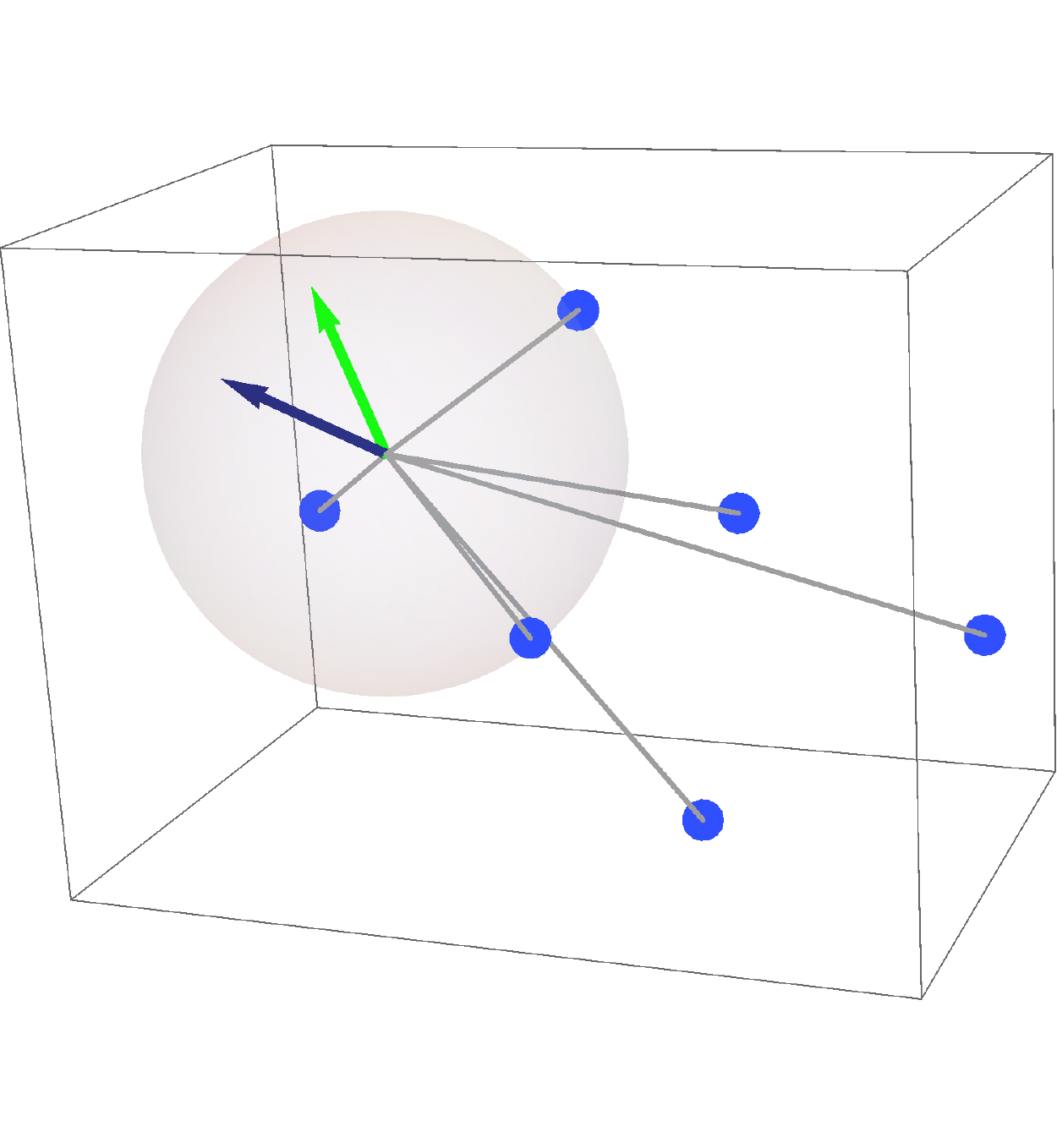}
		\label{fig:NormalizationBroken}
	}
	\caption{Bloch sphere with vectors, corresponding to the measurements (light blue dots), the true state (green arrow), and the current estimator (blue arrow). The initial measurement set and corresponding vectors are perfectly normalized (Fig.~\ref{fig:NormalizationOk}). After the transformation~\eqref{eq:MTransform} this normalization is broken (Fig.~\ref{fig:NormalizationBroken}). Thus, a consequent manual normalization is required, and the unnormalized vector length is interpreted as a measurement exposition time.}
	\label{fig:NormalizationViolation}
\end{figure}

Broken POVM normalization is not the only unwanted effect of the measurement transformation. Even with complemented $M_\gamma^\text{new}$, the normalization of each element may be violated:
\begin{equation}
\Tr (M_\gamma^\text{new}) \neq 1.
\end{equation}
If we put the vectors, corresponding to~$M_\gamma$, on a Bloch sphere, then generally after the transformation, these vectors will no more point to the surface of the sphere, as shown in Fig.~\ref{fig:NormalizationViolation}~\footnote{Strictly speaking, the Bloch sphere depicts only the normalized matrices~$M$ with $\Tr M = 1$, and the space outside the sphere corresponds to negative-definite matrices. However, in Fig.~\ref{fig:NormalizationViolation}, we treat the definition of the Bloch sphere rather freely, meaning that the operator $M$ and the corresponding vector~$s$ are tied by a relation: $M = \frac{\Tr M +s \sigma}{2}$, where $\sigma$ is the vector of three Pauli matrices.}. 

However, the real experimental setup always performs measurements corresponding to the normalized projectors, so a manual rescaling of $M_\gamma^\text{new}$ is required. In order to keep the likelihood value~\eqref{eq:PoissonLikelihood} unaltered after the rescaling, we notice that the probabilities~$p_j$ are always multiplied by the exposition times~$t_j$. Therefore, the value of $\Tr M_\gamma^\text{new}$ is interpreted as an exposition time~$t_\gamma$ of the corresponding normalized measurement.

\begin{figure*}
	\centering
	\subfloat[Initial view.]
	{
		\includegraphics[width=0.24\linewidth]{Iteration0}
		\label{fig:Iteration0}
	}
	\subfloat[The first iteration.]
	{
		\includegraphics[width=0.24\linewidth]{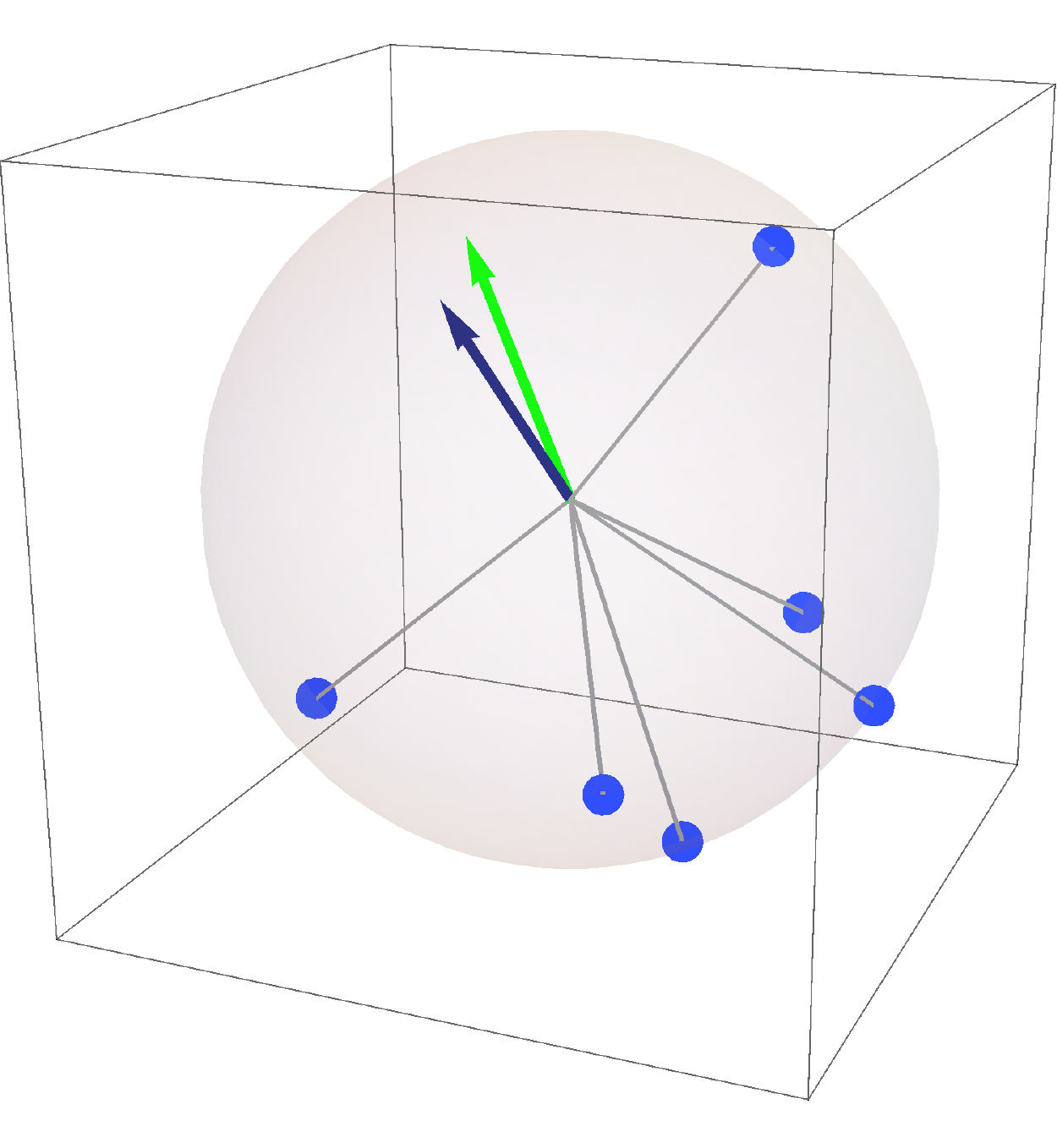}
		\label{fig:OneIter}
	}
	\subfloat[Ten iterations.]
	{
		\includegraphics[width=0.24\linewidth]{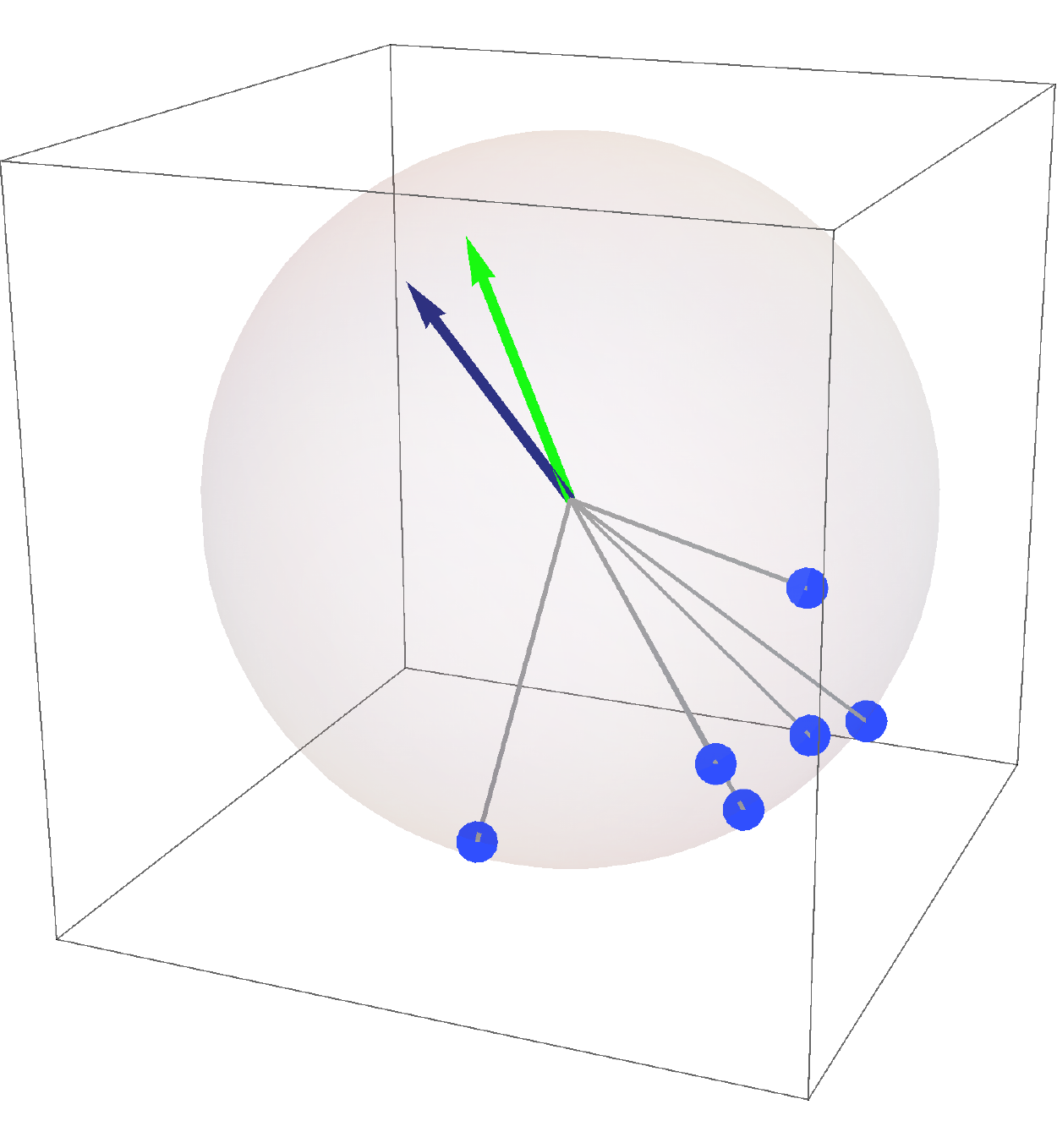}
		\label{fig:TwoIter}
	}
	\subfloat[Fifty iterations.]
	{
		\includegraphics[width=0.24\linewidth]{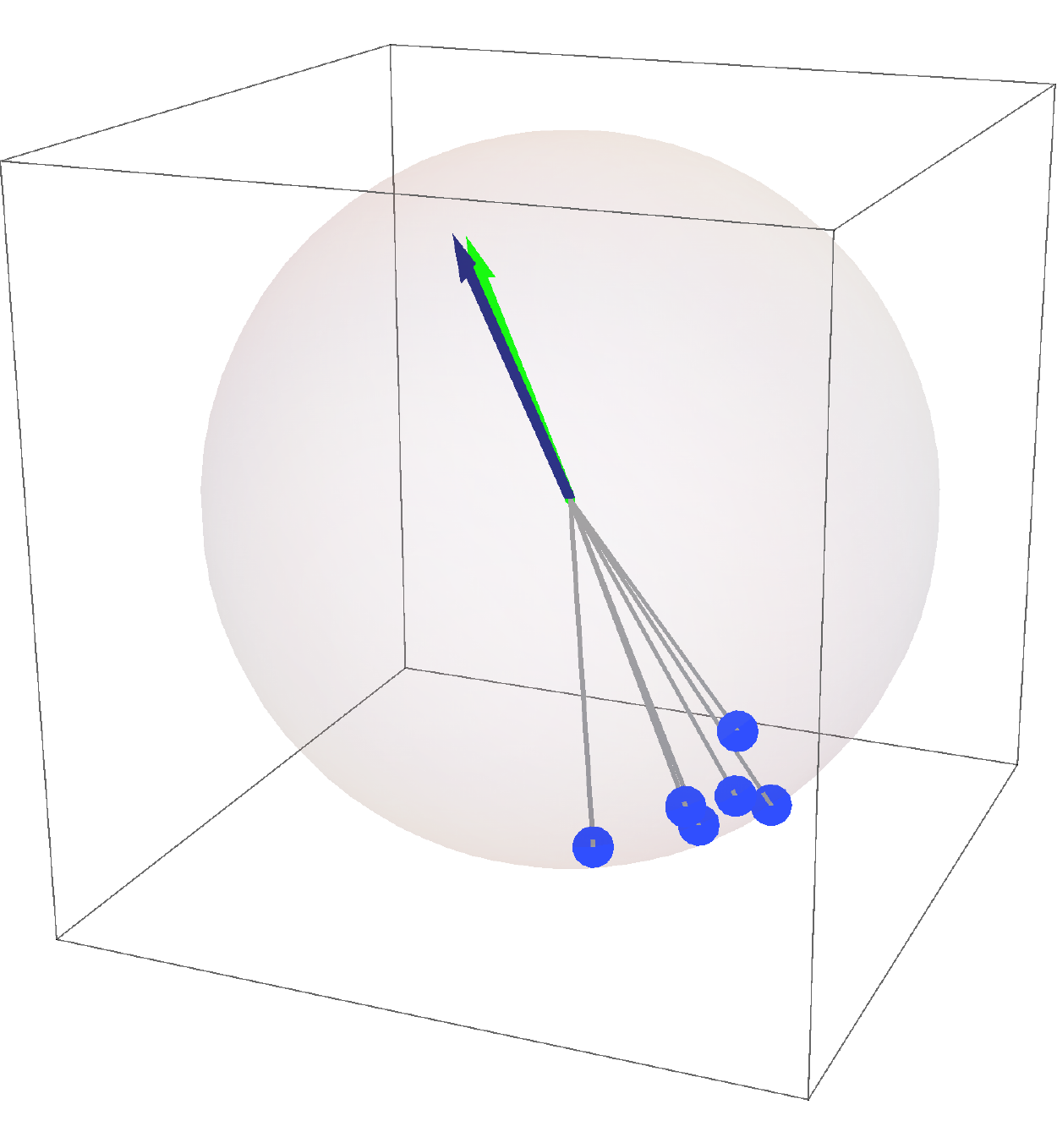}
		\label{fig:ThreeIter}
	}
	\caption{View of the Bloch sphere with vectors, corresponding to the normalized measurements (light blue dots), the true state (green arrow), and the current estimator (blue arrow). The initial measurement set is composed of three mutually unbiased bases. A priori estimator is the fully mixed state (the center of the Bloch sphere). With increasing iteration number, the measurements localize near the direction orthogonal to the true state (but not exactly on it, if the true state is not pure).}
	\label{fig:Transformations}
\end{figure*}

An example of the evolution of the normalized measurement vectors, while the estimator~$\hat \rho$ refines, is shown in Fig.~\ref{fig:Transformations}. As one can see, with increasing number of iterations the vectors move closer and closer to the direction, which is orthogonal to the true state. At the same time, the length of the corresponding unnormalized vectors (not shown in the figure) becomes larger, meaning that the required exposition time increases.

\section{Simulations and Experiment\label{sec:SimAndExp}}
\subsection{Numerical simulations}
We performed numerical simulations of single-qubit tomography with the rank-preserving-transformations protocol (RankP), random measurement strategy (Random), and measurements in the eigenbasis of the current estimator (Eigen)~\cite{Steinberg_PRL13,Kulik_PRA18}. We emphasize that RankP protocol requires a selection of some ``underlying'' set of measurements, subject to the transformation. We have chosen these measurements to be projectors corresponding to the vectors of mutually unbiased bases (MUB)~\cite{Wootters89}. For a physical realization of qubits as polarization states of single photons the corresponding vectors are
\begin{equation}
\{H, V, D, A, R, L\},
\end{equation}
where $\{H, V\}$ stands for the horizontal and vertical polarization, respectively, $\{D, A\}$ are $\pm 45^\circ$-linearly polarized states, $\{R, L\}$ are right and left circularly polarized states. A distinguishing feature of MUB is that the measurement performed in one basis does not provide any information about the measurement outcomes in any other basis. Measurement in MUB maximizes the information extracted about the unknown state on the first iteration and provides a good starting point for further adaptive algorithm operation.

\begin{figure*}
	\centering
	\subfloat[Random pure states.]
	{
		\includegraphics[width=0.49\linewidth]{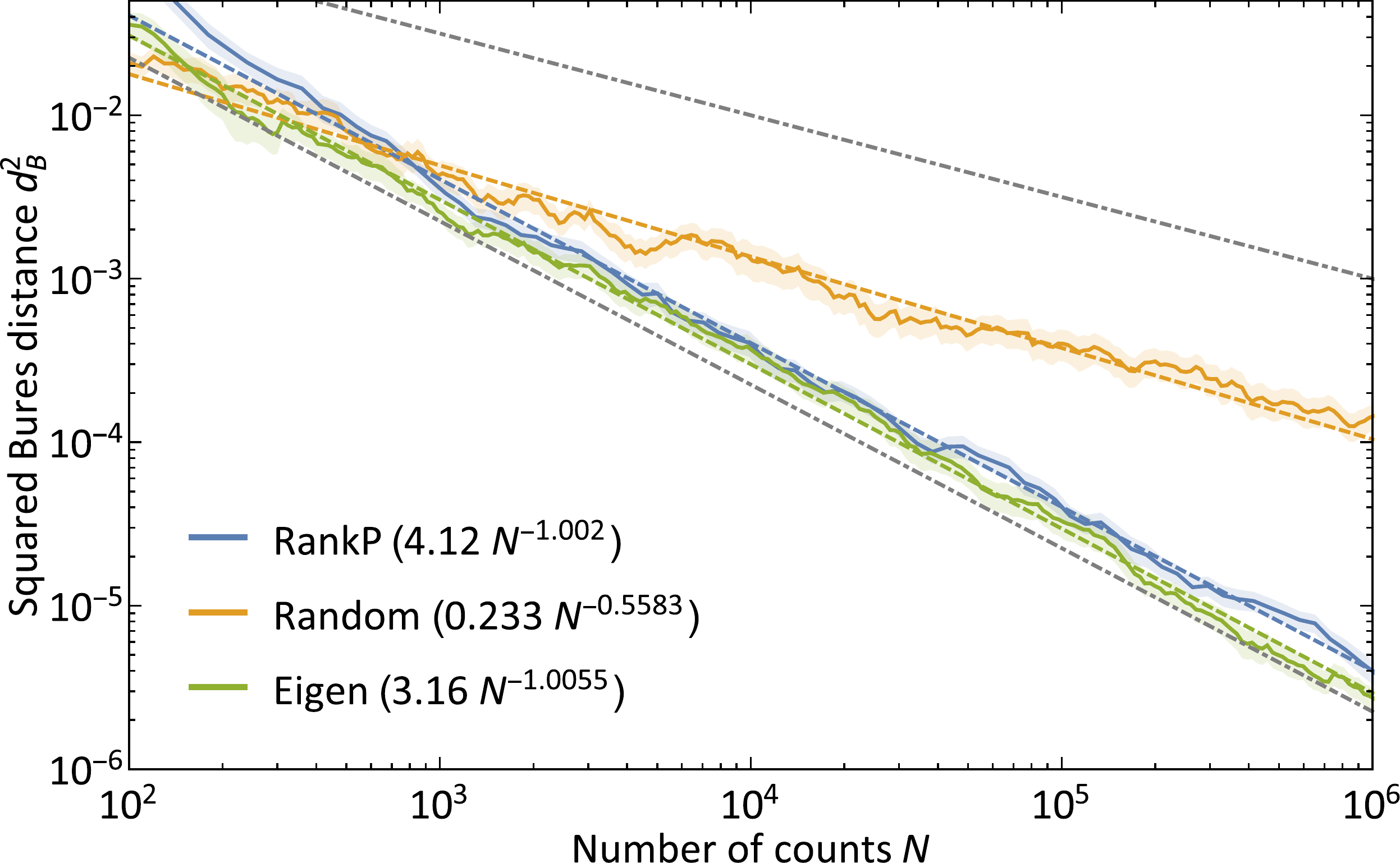}
		\label{fig:SimPure}
	}
	\subfloat[Random Bures-distributed states.]
	{
		\includegraphics[width=0.49\linewidth]{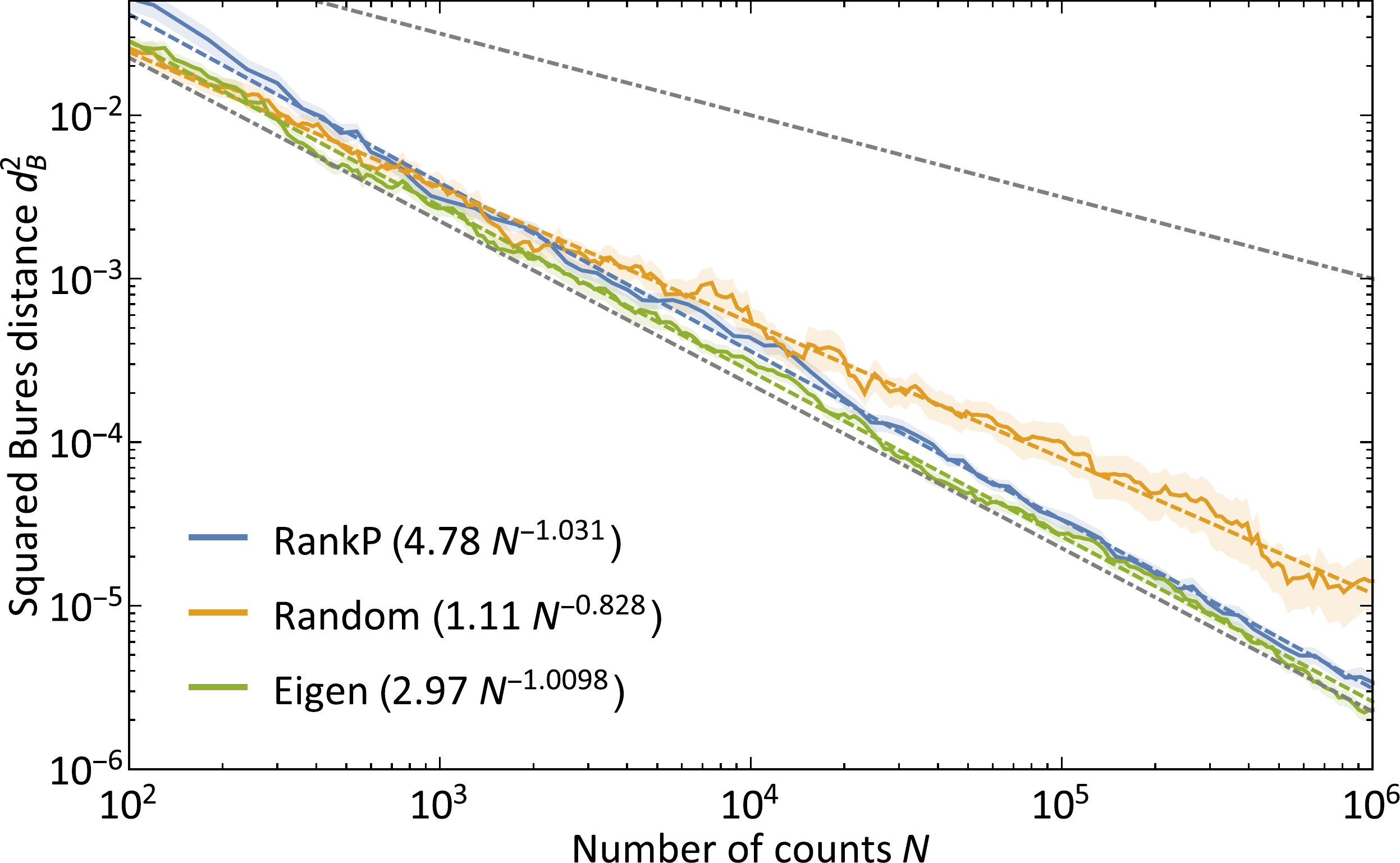}
		\label{fig:SimBures}
	}
	\caption{Averaged results for numerical simulations of (a) random pure and (b) Bures-distributed state tomography. All dependencies are averaged over 50 tomography runs. Here and below, gray dot-dashed lines depict dependencies $9/(4N)$ (lower) and $1 / \sqrt N$ (upper). The former corresponds to the asymptotic Gill-Massar bound~\cite{Massar_PRA00}, while the latter is given for illustrative purposes to show a typical slope of the dependencies for nonadaptive protocols, when recovering nearly pure states. Colored dashed lines show approximations of obtained dependencies with the power-law function (see legends for the best-fit parameters). Shaded regions show one standard deviation of the mean.}
	\label{fig:Simulations}
\end{figure*}

Simulation results are presented in Fig.~\ref{fig:Simulations}, where averaged dependencies of a qubit-tomography error on the number~$N$ of input state copies are demonstrated. As mentioned in the section~\ref{sec:Complementation}, the POVM normalization requirement~\eqref{eq:POVMNormalizationViolance} is violated for the transformed measurements~$M_\gamma^\text{new}$. Hence, the number~$N_\text{det}$ of \emph{detected} photons differs from the amount~$N_\text{emit}$ of \emph{emitted} photons from the source (not all photons are registered even in the case of unity detection efficiency). While $N_\text{det}$ is a directly measurable quantity, $N_\text{emit}$ should be used for protocol assessment and comparison with theoretical bounds. We estimate~$N_\text{emit}$ by multiplying the total exposition time of the experiment by the source intensity~$I$, which was measured prior to carrying out tomography. Here and below, the value of~$N\equiv N_\text{emit}$ corresponds to this estimation.

The state estimation error is determined as the squared Bures distance $d_B^2(\rho, \hat \rho)$ between the true state~$\rho$ and the current estimator~$\hat \rho$. Averaging is performed over 50 dependencies. For each tomography run, a new random pure state is drawn from a distribution that is uniform with respect to Haar measure (Fig.~\ref{fig:SimPure}). The colored dashed lines show approximations of the dependencies by a power function $d_B^2(N) = \alpha N^\beta$. Fig.~\ref{fig:SimBures} demonstrates analogous dependencies, but the true states are mixed and are drawn uniformly w.r.t. the measure, induced by the Bures distance~\cite{Zyczkowski_JMP11}.

An asymptotic behavior of the dependencies $d_B^2(N)$ was studied for different cases in the works~\cite{Massar_PRA00,Bagan_PRL2006,Bogdanov_PRA2011}. For qubit tomography, where measurements form a decomposition of unity, the asymptotic limit is $9/(4N)$ for mixed states and $1/N$---for pure. However, static protocols are unable to reach this limit for (nearly) pure states and demonstrate only $\propto 1/\sqrt{N}$ convergence. The $1/N$ limit becomes achievable only with adaptive tomography algorithms.


As one can see from the best-fit parameters, the RankP method demonstrates a fast convergence $\propto 1/N$, as expected for an adaptive tomography method. Moreover, it almost achieves the theoretical threshold $9/(4N)$, certifying that the rank-preserving-transformations algorithm is a highly efficient tomography method. Nevertheless, eigenbasis tomography demonstrates a slight but noticeable superiority. 

\subsection{Experiment}

\begin{figure}
	\centering
	\includegraphics[width=1\linewidth]{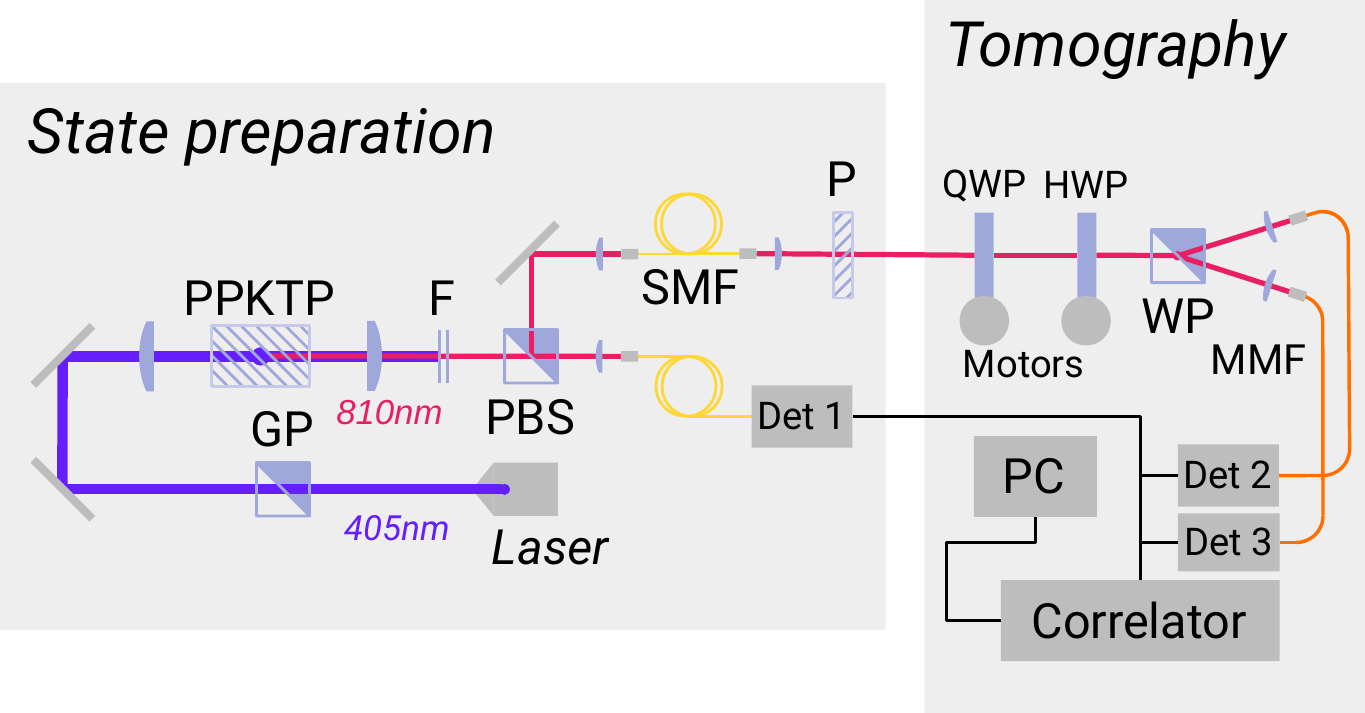}
	\caption{Experimental setup. A diode laser pumps a nonlinear crystal PPKTP, where spontaneous parametric down-conversion occurs. One photon from the pair serves as a trigger, while another one is subject to the tomographic procedure. Polarizer~P prepares the true state, while waveplates QWP and HWP together with a Wollaston prism (WP) perform the desired measurement.}
	\label{fig:Setup}
\end{figure}

We compared the protocols in an experiment with polarization qubits, produced by a heralded single-photon source. The experimental setup scheme is presented in Fig.~\ref{fig:Setup}. A nonlinear crystal (PPKTP) is pumped by a 405-nm diode laser. Photon pairs are generated in a type-II collinear spontaneous parametric down-conversion (SPDC) process. Pump polarization is determined by a Glan prism (GP). The pump propagated through the PPKTP crystal is blocked by a long-pass filter with a cut-off wavelength of 450~nm and a narrow-band interference filter (F) with transmitting band of $810 \pm 5$~nm (full width at half maximum). After a polarizing beam splitter (PBS), the photons are coupled into single-mode optical fibers (SMF). One photon from the pair goes directly to the detector acting as a trigger, and another one is subjected to the tomographic procedure. The measurement projection is determined by a thin-film polarizer (P) with an extinction coefficient of at least $10^5$. Measurements are performed by a quarter- (QWP) and a half-waveplate (HWP) followed by a Wollaston prism (WP). Orientations of the plates axes are set by motors, controlled from a computer (PC), with an accuracy of at least $0.1^\circ$. The two output channels of the Wollaston prism are coupled to two separate multimode optical fibers (MMF), connected to single-photon detectors. Coincidences between the signal and idler (trigger) arms are processed by the start-stop module, which sends data to a PC. 

\begin{figure}
	\centering
	\includegraphics[width=1\linewidth]{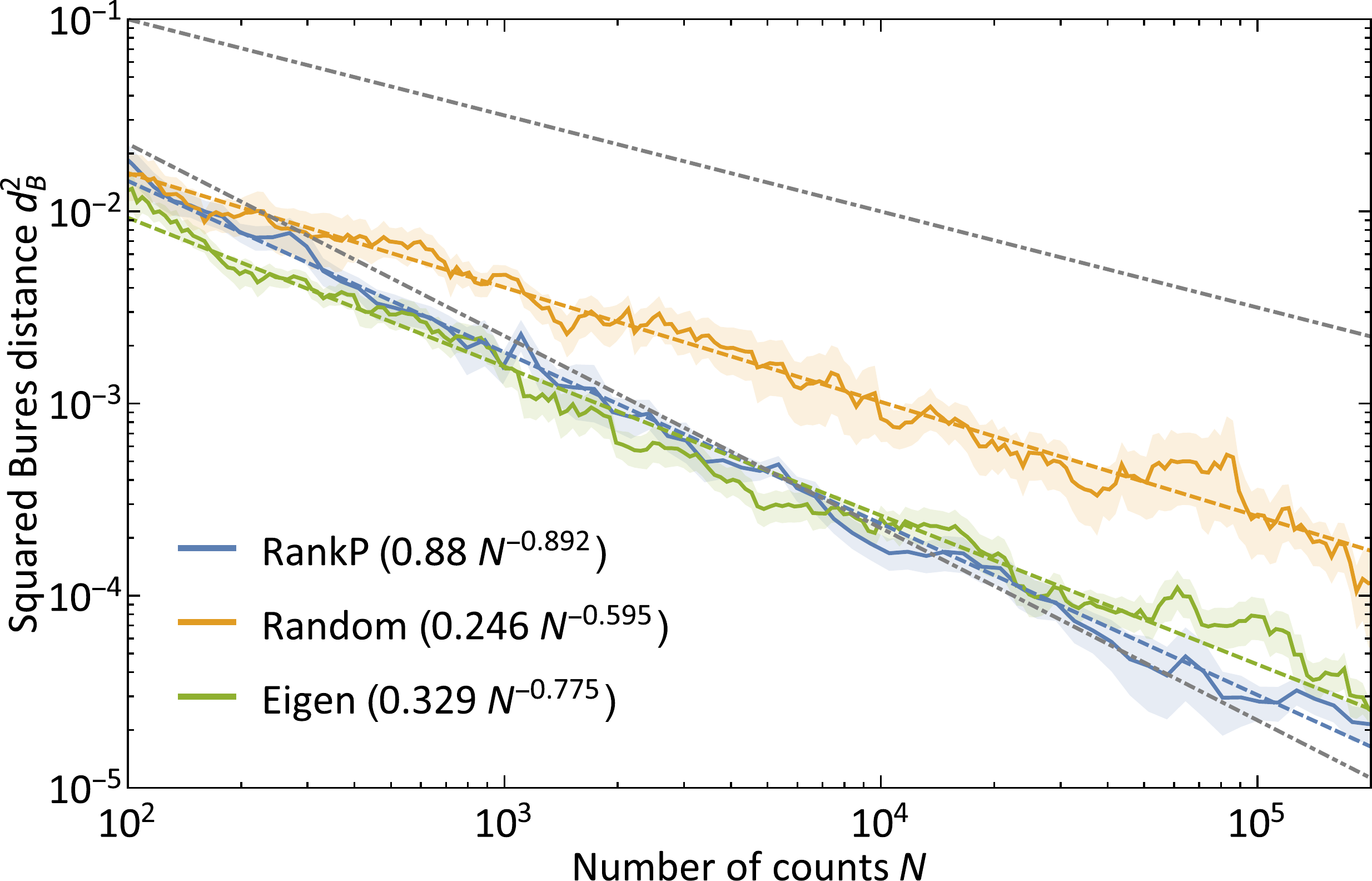}
	\caption{Experimentally obtained dependencies of the squared Bures distance $d_B^2(N)$ between the current estimator~$\hat \rho(N)$ and the final assessment~$\hat \rho(N_0)$ for $N_0 = 5 \times 10^5$ on the number of input state copies~$N$. Curves are averaged over 14--19 tomography runs (depending on the protocol).}
	\label{fig:Experiment}
\end{figure}

Fig.~\ref{fig:Experiment} demonstrates the experimentally observed precision of the tomographic estimate averaged over 14--19 experimental dependencies for different protocols. Note that we cannot know the true measured state in the real experiment, so we use the distance from the current estimator to the final estimate obtained for the number of counts $N_0 \approx 5 \times 10^5$ as a measure of precision. Therefore, the dependencies $d^2_B(N)$ will always end up with a sharp decline towards zero, and Fig.~\ref{fig:Experiment} is clipped before this decline occurs.

The behavior is almost similar to the simulated results---Eigen and RankP protocols are both very close to the theoretical bound $9/(4N)$. But now, according to the fit results, RankP gets a slight convergence advantage out of a statistical error range. Despite that, we do not conclude that RankP or Eigen achieve any noticeable superiority in this experiment. Both algorithms demonstrate nearly equal efficiency in absolute values of the Bures distance~$d_B^2$.

An important note about all the presented results is that our experimental setup allows us to carry out two orthogonal measurements constituting some basis at once~\footnote{An ability to simultaneously measure two orthogonal projectors was also taken into account in the simulations.}. Consequently, eigen- and random-basis strategies proceed twice faster than the RankP method, where the projectors are not orthogonal, thus they are forced to be measured one-by-one (the registered counts from the third detector in Fig.~\ref{fig:Setup} are neglected). However, this drawback can be circumvented if an experimental setup can simultaneously measure the whole set of nonorthogonal projectors, making the RankP protocol advantageous for such setups. As an example of the nonorthogonal POVM implementation, we mention Ref.~\cite{Ling06}, where a tetrahedron qubit POVM was realized. We believe the ideas from this work can be adopted for constructing more complex POVMs suitable for the realization of the RankP protocol. 

\begin{figure*}
	\centering
	\subfloat[Random pure states.]
	{
		\includegraphics[width=0.49\linewidth]{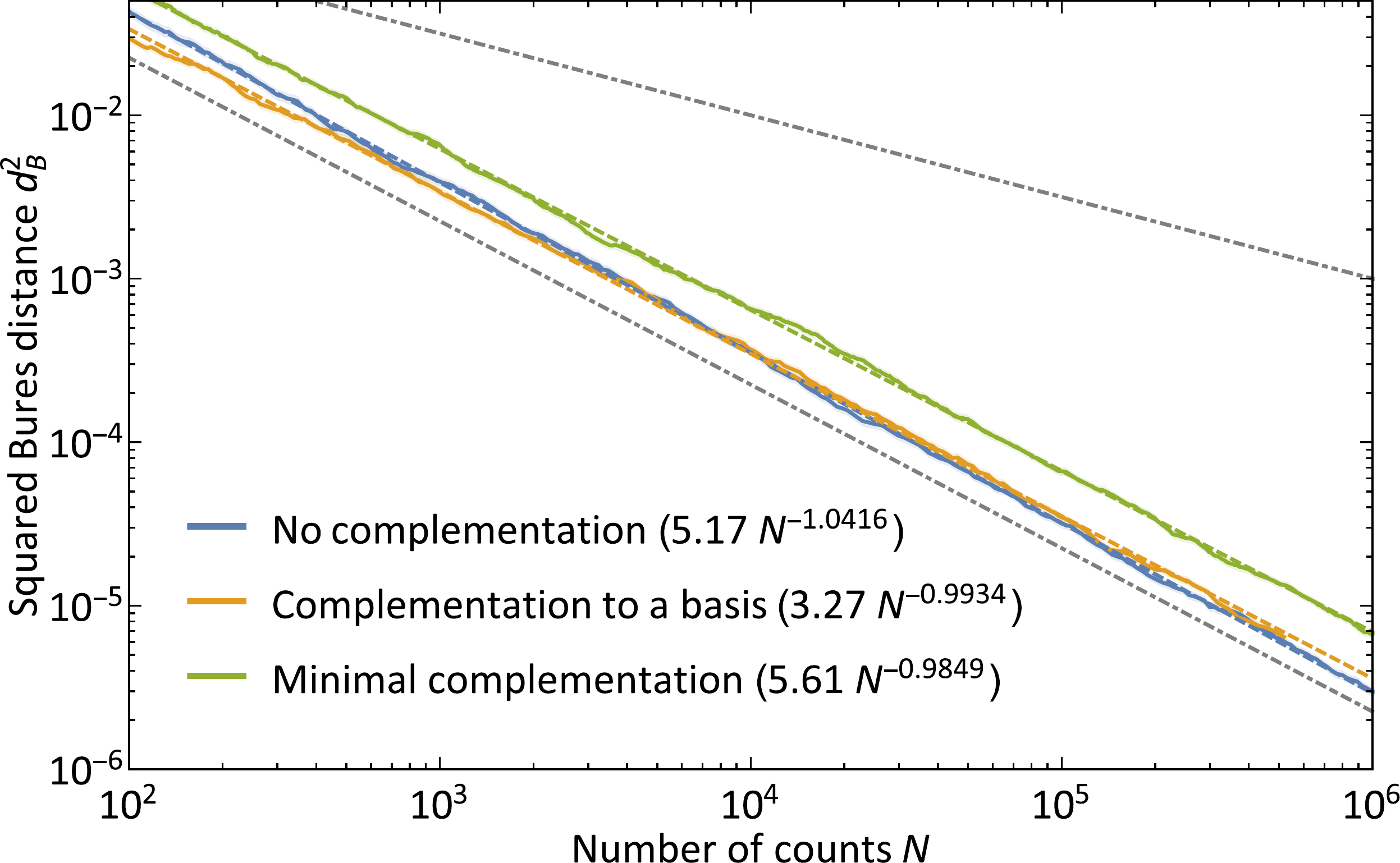}
		\label{fig:Purecomplementation}
	}
	\subfloat[Random Bures-distributed states.]
	{
		\includegraphics[width=0.49\linewidth]{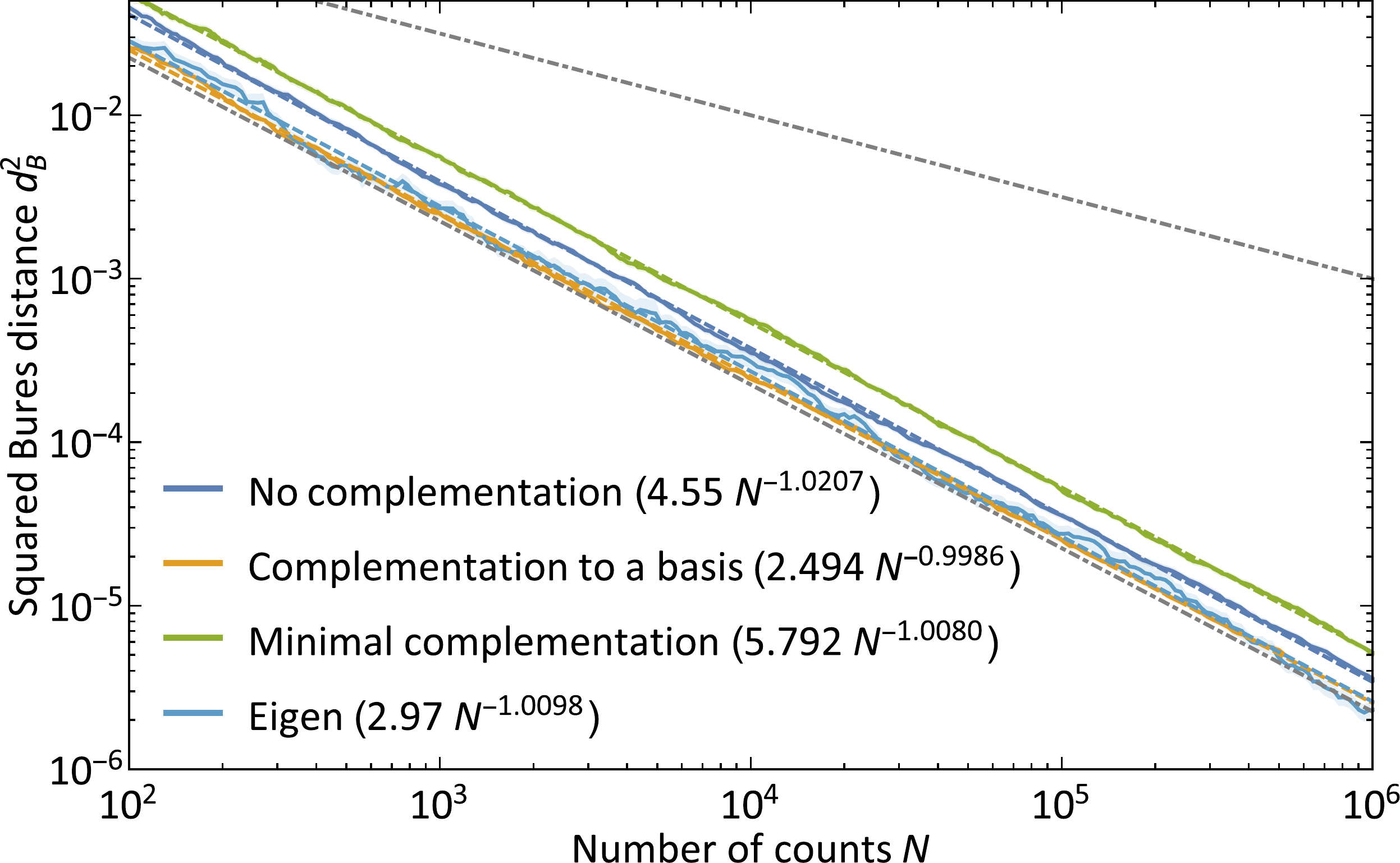}
		\label{fig:BuresComplementation}
	}
	\caption{Simulated results for different measurement complementation schemes. Different complementation strategies have identical convergence rate, but different constant prefactors (see text for details).}
	\label{fig:Complementation}
\end{figure*}

\subsection{Complemented measurements}
An important question is whether it is possible to improve convergence speed and accuracy of quantum state tomography with the rank-preserving-transformations algorithm by complementing the measurement set as described in the section~\ref{sec:Complementation} or by using unitary freedom~\eqref{eq:UnitaryFreedom}. Simulations with different unitary rotations~$V$ applied to the transformation~$\mathcal L$ revealed no significant influence out of the statistical error range. However, simulations of measurement complementation lead to more exciting results that are presented in Fig.~\ref{fig:Complementation}. In the case of pure state tomography (Fig.~\ref{fig:Purecomplementation}), complementation to a basis (RankP-B) gives no difference. As long as the setup is assumed to perform a pair of orthogonal projectors at once, the absence of difference between original, no-complemented RankP (RankP-NC) and RankP-B can only mean that no useful information about the unknown state is extracted from additional measurements. 

At the same time, tomography with RankP-M modification is slower than RankP-NC by a constant factor. The reason is that according to the formulae~\eqref{eq:ComplementationExplained} and~\eqref{eq:Complementation}, actually, only one additional measurement is sufficient to complement the set (the second measurement has a vanishing exposition time: $\lambda_2 = 0$). During adaptive tomography iterations, the measurements are localized close to the direction pointed orthogonally to the current estimator (see Fig.~\ref{fig:Transformations}). Consequently, in the limit of large~$N$, all six non-complemented measurements are almost co-directional, and the exposition time of the only additional measurement will be about the total exposition of all non-complemented measurements. As a result, after the complementation, the total time approximately doubles.

To estimate the averaged ratio~$\bar R$ of accuracies for two tomography methods on some interval $[N_1, N_2]$, we can calculate a geometric mean of these ratios at the edges of this interval from the best fit parameters $\alpha_1$, $\alpha_2$, $\beta_1$, $\beta_2$ (recall that the fit formula is $d_B^2(N) = \alpha N^\beta$):
\begin{equation}
\bar R =  \frac{\alpha_2}{\alpha_1}(N_1 N_2)^\frac{\beta2 - \beta1}{2}.
\label{eq:EfficiencyRatio}
\end{equation}

According to Eq.~\eqref{eq:EfficiencyRatio}, the average ratio between RankP-M and RankP-NC tomography is $\bar R = 1.83$ for $N \in [10^2, 10^6]$. This value is less than the anticipated one, $R = 2$. This behavior can be connected with the fact that the measurement vectors never become precisely orthogonal to the current estimator. The ratio between RankP-B and RankP-NC is $\bar R = 0.986 \approx 1$ as expected.

For Bures-random mixed states, complementation to a basis becomes the most effective strategy (Fig.~\ref{fig:BuresComplementation}). It means that orthogonal measurements can give more useful information about the unknown quantum state if it is mixed. Considering that the experimental setup measures an orthogonal basis as a whole, complementation to a basis is expected to be the most effective for general use. Compared to Eigen protocol, both methods demonstrate fast convergence that is close to the Gill-Massar bound. Averaged accuracy ratio between RankP-M and RankP-NC is $\bar R = 1.43$. The most important ratio between RankP-B and Eigen is  $\bar R = 1.07$. It indicates that none of these methods prevail, i.\,e., they both are highly efficient in general.

\section{Conclusion}

We have experimentally tested a novel approach to adaptive quantum tomography based on rank-preserving transformations and compared it to the known close-to-optimal protocols based on measurements in the estimator eigenbasis. Our results, both in numerical simulations and with real experimental data, suggest that there is almost no significant difference between the two protocols.

The reason why we do not observe any super-efficiency for the RankP protocol as compared to Ref.~\cite{Bogdanov_SPIE2019} lies in the way we treat~$N$. In our approach $N=N_\text{emit}$ is the \emph{number of copies} of the quantum state consumed in the protocol, in full accordance with the setting for which the information-theoretical bounds are derived~\cite{Massar_PRA00}. At the same time, in Ref.~\cite{Bogdanov_SPIE2019} $N=N_\text{det}$ is the \emph{number of detected events}, and may be significantly lower (especially for near-pure states) because the result of the rank-preserving transformation is not a decomposition of unity. Physically, this scenario corresponds to an incomplete measurement, where some of the outcomes are not registered. If we complement the set of measurements with those corresponding to the unregistered outcomes, we arrive at a complete (although non-orthogonal) POVM, and all the analysis presented here becomes valid. As we have shown, various complementation strategies have only slightly different effects on the statistical efficiency of the protocol. 

Finally, let us note that due to technical restrictions of our experimental setup we were only able to implement orthogonal projective measurements, introducing additional overhead. However, even the ability to implement arbitrary POVM in a single-shot manner will not provide efficiency beyond known limits if the protocol is treated as described above.

\begin{acknowledgments}
	We acknowledge financial support from Russian Foundation for Basic Research (RFBR Project No. 19-32-80043 and RFBR Project No. 19-52-80034), Fund for Assistance to Small Innovative Enterprises (FASIE Contract No. 15557GU/2020), and support under the Russian National Technological Initiative via MSU Quantum Technology Centre.
\end{acknowledgments}

\bibliographystyle{apsrev4-1}
\bibliography{ref_base}

\begin{thebibliography}{25}%
\makeatletter
\providecommand \@ifxundefined [1]{%
 \@ifx{#1\undefined}
}%
\providecommand \@ifnum [1]{%
 \ifnum #1\expandafter \@firstoftwo
 \else \expandafter \@secondoftwo
 \fi
}%
\providecommand \@ifx [1]{%
 \ifx #1\expandafter \@firstoftwo
 \else \expandafter \@secondoftwo
 \fi
}%
\providecommand \natexlab [1]{#1}%
\providecommand \enquote  [1]{``#1''}%
\providecommand \bibnamefont  [1]{#1}%
\providecommand \bibfnamefont [1]{#1}%
\providecommand \citenamefont [1]{#1}%
\providecommand \href@noop [0]{\@secondoftwo}%
\providecommand \href [0]{\begingroup \@sanitize@url \@href}%
\providecommand \@href[1]{\@@startlink{#1}\@@href}%
\providecommand \@@href[1]{\endgroup#1\@@endlink}%
\providecommand \@sanitize@url [0]{\catcode `\\12\catcode `\$12\catcode
  `\&12\catcode `\#12\catcode `\^12\catcode `\_12\catcode `\%12\relax}%
\providecommand \@@startlink[1]{}%
\providecommand \@@endlink[0]{}%
\providecommand \url  [0]{\begingroup\@sanitize@url \@url }%
\providecommand \@url [1]{\endgroup\@href {#1}{\urlprefix }}%
\providecommand \urlprefix  [0]{URL }%
\providecommand \Eprint [0]{\href }%
\providecommand \doibase [0]{http://dx.doi.org/}%
\providecommand \selectlanguage [0]{\@gobble}%
\providecommand \bibinfo  [0]{\@secondoftwo}%
\providecommand \bibfield  [0]{\@secondoftwo}%
\providecommand \translation [1]{[#1]}%
\providecommand \BibitemOpen [0]{}%
\providecommand \bibitemStop [0]{}%
\providecommand \bibitemNoStop [0]{.\EOS\space}%
\providecommand \EOS [0]{\spacefactor3000\relax}%
\providecommand \BibitemShut  [1]{\csname bibitem#1\endcsname}%
\let\auto@bib@innerbib\@empty
\bibitem [{\citenamefont {Paris}\ and\ \citenamefont {\ifmmode \check{R}\else
  \v{R}\fi{}eh\'a\ifmmode~\check{c}\else \v{c}\fi{}ek}(2004)}]{Paris_Book2004}%
  \BibitemOpen
  \bibinfo {editor} {\bibfnamefont {M.}~\bibnamefont {Paris}}\ and\ \bibinfo
  {editor} {\bibfnamefont {J.}~\bibnamefont {\ifmmode \check{R}\else
  \v{R}\fi{}eh\'a\ifmmode~\check{c}\else \v{c}\fi{}ek}},\ eds.,\ \href
  {\doibase 10.1007/b98673} {\emph {\bibinfo {title} {Quantum State
  Estimation}}},\ \bibinfo {series} {Lecture Notes in Physics}, Vol.\ \bibinfo
  {volume} {649}\ (\bibinfo  {publisher} {Springer-Verlag},\ \bibinfo {year}
  {2004})\BibitemShut {NoStop}%
\bibitem [{\citenamefont {Bogdanov}\ \emph {et~al.}(2010)\citenamefont
  {Bogdanov}, \citenamefont {Brida}, \citenamefont {Genovese}, \citenamefont
  {Kulik}, \citenamefont {Moreva},\ and\ \citenamefont
  {Shurupov}}]{Kulik_PRL10}%
  \BibitemOpen
  \bibfield  {author} {\bibinfo {author} {\bibfnamefont {Y.~I.}\ \bibnamefont
  {Bogdanov}}, \bibinfo {author} {\bibfnamefont {G.}~\bibnamefont {Brida}},
  \bibinfo {author} {\bibfnamefont {M.}~\bibnamefont {Genovese}}, \bibinfo
  {author} {\bibfnamefont {S.~P.}\ \bibnamefont {Kulik}}, \bibinfo {author}
  {\bibfnamefont {E.~V.}\ \bibnamefont {Moreva}}, \ and\ \bibinfo {author}
  {\bibfnamefont {A.~P.}\ \bibnamefont {Shurupov}},\ }\href {\doibase
  10.1103/PhysRevLett.105.010404} {\bibfield  {journal} {\bibinfo  {journal}
  {Phys. Rev. Lett.}\ }\textbf {\bibinfo {volume} {105}},\ \bibinfo {pages}
  {010404} (\bibinfo {year} {2010})}\BibitemShut {NoStop}%
\bibitem [{\citenamefont {Gill}\ and\ \citenamefont
  {Massar}(2000)}]{Massar_PRA00}%
  \BibitemOpen
  \bibfield  {author} {\bibinfo {author} {\bibfnamefont {R.~D.}\ \bibnamefont
  {Gill}}\ and\ \bibinfo {author} {\bibfnamefont {S.}~\bibnamefont {Massar}},\
  }\href {\doibase 10.1103/PhysRevA.61.042312} {\bibfield  {journal} {\bibinfo
  {journal} {Phys. Rev. A}\ }\textbf {\bibinfo {volume} {61}},\ \bibinfo
  {pages} {042312} (\bibinfo {year} {2000})}\BibitemShut {NoStop}%
\bibitem [{\citenamefont {Bagan}\ \emph {et~al.}(2006)\citenamefont {Bagan},
  \citenamefont {Ballester}, \citenamefont {Gill}, \citenamefont
  {Mu{\~n}oz-Tapia},\ and\ \citenamefont {Romero-Isart}}]{Bagan_PRL2006}%
  \BibitemOpen
  \bibfield  {author} {\bibinfo {author} {\bibfnamefont {E.}~\bibnamefont
  {Bagan}}, \bibinfo {author} {\bibfnamefont {M.~A.}\ \bibnamefont
  {Ballester}}, \bibinfo {author} {\bibfnamefont {R.~D.}\ \bibnamefont {Gill}},
  \bibinfo {author} {\bibfnamefont {R.}~\bibnamefont {Mu{\~n}oz-Tapia}}, \ and\
  \bibinfo {author} {\bibfnamefont {O.}~\bibnamefont {Romero-Isart}},\ }\href
  {\doibase 10.1103/PhysRevLett.97.130501} {\bibfield  {journal} {\bibinfo
  {journal} {Phys. Rev. Lett.}\ }\textbf {\bibinfo {volume} {97}},\ \bibinfo
  {pages} {130501} (\bibinfo {year} {2006})}\BibitemShut {NoStop}%
\bibitem [{\citenamefont {Straupe}(2016)}]{Straupe_JETP2016}%
  \BibitemOpen
  \bibfield  {author} {\bibinfo {author} {\bibfnamefont {S.~S.}\ \bibnamefont
  {Straupe}},\ }\href@noop {} {\bibfield  {journal} {\bibinfo  {journal} {JETP
  letters}\ }\textbf {\bibinfo {volume} {104}},\ \bibinfo {pages} {510}
  (\bibinfo {year} {2016})}\BibitemShut {NoStop}%
\bibitem [{\citenamefont {Husz{\'a}r}\ and\ \citenamefont
  {Houlsby}(2012)}]{Houlsby_PRA12}%
  \BibitemOpen
  \bibfield  {author} {\bibinfo {author} {\bibfnamefont {F.}~\bibnamefont
  {Husz{\'a}r}}\ and\ \bibinfo {author} {\bibfnamefont {N.~M.~T.}\ \bibnamefont
  {Houlsby}},\ }\href {\doibase 10.1103/PhysRevA.85.052120} {\bibfield
  {journal} {\bibinfo  {journal} {Phys. Rev. A}\ }\textbf {\bibinfo {volume}
  {85}},\ \bibinfo {pages} {052120} (\bibinfo {year} {2012})}\BibitemShut
  {NoStop}%
\bibitem [{\citenamefont {Mahler}\ \emph {et~al.}(2013)\citenamefont {Mahler},
  \citenamefont {Rozema}, \citenamefont {Darabi}, \citenamefont {Ferrie},
  \citenamefont {Blume-Kohout},\ and\ \citenamefont
  {Steinberg}}]{Steinberg_PRL13}%
  \BibitemOpen
  \bibfield  {author} {\bibinfo {author} {\bibfnamefont {D.~H.}\ \bibnamefont
  {Mahler}}, \bibinfo {author} {\bibfnamefont {L.~A.}\ \bibnamefont {Rozema}},
  \bibinfo {author} {\bibfnamefont {A.}~\bibnamefont {Darabi}}, \bibinfo
  {author} {\bibfnamefont {C.}~\bibnamefont {Ferrie}}, \bibinfo {author}
  {\bibfnamefont {R.}~\bibnamefont {Blume-Kohout}}, \ and\ \bibinfo {author}
  {\bibfnamefont {A.~M.}\ \bibnamefont {Steinberg}},\ }\href {\doibase
  10.1103/PhysRevLett.111.183601} {\bibfield  {journal} {\bibinfo  {journal}
  {Phys. Rev. Lett.}\ }\textbf {\bibinfo {volume} {111}},\ \bibinfo {pages}
  {183601} (\bibinfo {year} {2013})}\BibitemShut {NoStop}%
\bibitem [{\citenamefont {Hou}\ \emph {et~al.}(2016)\citenamefont {Hou},
  \citenamefont {Zhu}, \citenamefont {Xiang}, \citenamefont {Li},\ and\
  \citenamefont {Guo}}]{Guo_NPJQI16}%
  \BibitemOpen
  \bibfield  {author} {\bibinfo {author} {\bibfnamefont {Z.}~\bibnamefont
  {Hou}}, \bibinfo {author} {\bibfnamefont {H.}~\bibnamefont {Zhu}}, \bibinfo
  {author} {\bibfnamefont {G.-Y.}\ \bibnamefont {Xiang}}, \bibinfo {author}
  {\bibfnamefont {C.-F.}\ \bibnamefont {Li}}, \ and\ \bibinfo {author}
  {\bibfnamefont {G.-C.}\ \bibnamefont {Guo}},\ }\href {\doibase
  10.1038/npjqi.2016.1} {\bibfield  {journal} {\bibinfo  {journal} {npj Quantum
  Information}\ }\textbf {\bibinfo {volume} {2}},\ \bibinfo {pages} {16001}
  (\bibinfo {year} {2016})}\BibitemShut {NoStop}%
\bibitem [{\citenamefont {Sugiyama}\ \emph {et~al.}(2012)\citenamefont
  {Sugiyama}, \citenamefont {Turner},\ and\ \citenamefont
  {Murao}}]{Murao_PRA12}%
  \BibitemOpen
  \bibfield  {author} {\bibinfo {author} {\bibfnamefont {T.}~\bibnamefont
  {Sugiyama}}, \bibinfo {author} {\bibfnamefont {P.~S.}\ \bibnamefont
  {Turner}}, \ and\ \bibinfo {author} {\bibfnamefont {M.}~\bibnamefont
  {Murao}},\ }\href {\doibase 10.1103/PhysRevA.85.052107} {\bibfield  {journal}
  {\bibinfo  {journal} {Phys. Rev. A}\ }\textbf {\bibinfo {volume} {85}},\
  \bibinfo {pages} {052107} (\bibinfo {year} {2012})}\BibitemShut {NoStop}%
\bibitem [{\citenamefont {Okamoto}\ \emph {et~al.}(2012)\citenamefont
  {Okamoto}, \citenamefont {Iefuji}, \citenamefont {Oyama}, \citenamefont
  {Yamagata}, \citenamefont {Imai}, \citenamefont {Fujiwara},\ and\
  \citenamefont {Takeuchi}}]{Takeuchi_PRL12}%
  \BibitemOpen
  \bibfield  {author} {\bibinfo {author} {\bibfnamefont {R.}~\bibnamefont
  {Okamoto}}, \bibinfo {author} {\bibfnamefont {M.}~\bibnamefont {Iefuji}},
  \bibinfo {author} {\bibfnamefont {S.}~\bibnamefont {Oyama}}, \bibinfo
  {author} {\bibfnamefont {K.}~\bibnamefont {Yamagata}}, \bibinfo {author}
  {\bibfnamefont {H.}~\bibnamefont {Imai}}, \bibinfo {author} {\bibfnamefont
  {A.}~\bibnamefont {Fujiwara}}, \ and\ \bibinfo {author} {\bibfnamefont
  {S.}~\bibnamefont {Takeuchi}},\ }\href {\doibase
  10.1103/PhysRevLett.109.130404} {\bibfield  {journal} {\bibinfo  {journal}
  {Phys. Rev. Lett.}\ }\textbf {\bibinfo {volume} {109}},\ \bibinfo {pages}
  {130404} (\bibinfo {year} {2012})}\BibitemShut {NoStop}%
\bibitem [{\citenamefont {Kalev}\ and\ \citenamefont {Hen}(2015)}]{Hen_NJP15}%
  \BibitemOpen
  \bibfield  {author} {\bibinfo {author} {\bibfnamefont {A.}~\bibnamefont
  {Kalev}}\ and\ \bibinfo {author} {\bibfnamefont {I.}~\bibnamefont {Hen}},\
  }\href@noop {} {\bibfield  {journal} {\bibinfo  {journal} {New J. Phys.}\
  }\textbf {\bibinfo {volume} {17}},\ \bibinfo {pages} {093008} (\bibinfo
  {year} {2015})}\BibitemShut {NoStop}%
\bibitem [{\citenamefont {Ferrie}(2014)}]{Ferrie_PRL14}%
  \BibitemOpen
  \bibfield  {author} {\bibinfo {author} {\bibfnamefont {C.}~\bibnamefont
  {Ferrie}},\ }\href {\doibase 10.1103/PhysRevLett.113.190404} {\bibfield
  {journal} {\bibinfo  {journal} {Phys. Rev. Lett.}\ }\textbf {\bibinfo
  {volume} {113}},\ \bibinfo {pages} {190404} (\bibinfo {year}
  {2014})}\BibitemShut {NoStop}%
\bibitem [{\citenamefont {Granade}\ \emph {et~al.}(2017)\citenamefont
  {Granade}, \citenamefont {Ferrie},\ and\ \citenamefont
  {Flammia}}]{Granade_NJP2017}%
  \BibitemOpen
  \bibfield  {author} {\bibinfo {author} {\bibfnamefont {C.}~\bibnamefont
  {Granade}}, \bibinfo {author} {\bibfnamefont {C.}~\bibnamefont {Ferrie}}, \
  and\ \bibinfo {author} {\bibfnamefont {S.~T.}\ \bibnamefont {Flammia}},\
  }\href {\doibase 10.1088/1367-2630/aa8fe6} {\bibfield  {journal} {\bibinfo
  {journal} {New Journal of Physics}\ }\textbf {\bibinfo {volume} {19}},\
  \bibinfo {pages} {113017} (\bibinfo {year} {2017})}\BibitemShut {NoStop}%
\bibitem [{\citenamefont {Bogdanov}\ \emph {et~al.}(2019)\citenamefont
  {Bogdanov}, \citenamefont {Bogdanova}, \citenamefont {Bantysh},\ and\
  \citenamefont {Kuznetsov}}]{Bogdanov_SPIE2019}%
  \BibitemOpen
  \bibfield  {author} {\bibinfo {author} {\bibfnamefont {Y.~I.}\ \bibnamefont
  {Bogdanov}}, \bibinfo {author} {\bibfnamefont {N.~A.}\ \bibnamefont
  {Bogdanova}}, \bibinfo {author} {\bibfnamefont {B.~I.}\ \bibnamefont
  {Bantysh}}, \ and\ \bibinfo {author} {\bibfnamefont {Y.~A.}\ \bibnamefont
  {Kuznetsov}},\ }in\ \href {\doibase 10.1117/12.2522078} {\emph {\bibinfo
  {booktitle} {International Conference on Micro-and Nano-Electronics 2018}}},\
  Vol.\ \bibinfo {volume} {11022}\ (\bibinfo {organization} {International
  Society for Optics and Photonics},\ \bibinfo {year} {2019})\ p.\ \bibinfo
  {pages} {110222O}\BibitemShut {NoStop}%
\bibitem [{\citenamefont {Hradil}(1997)}]{Hradil97}%
  \BibitemOpen
  \bibfield  {author} {\bibinfo {author} {\bibfnamefont {Z.}~\bibnamefont
  {Hradil}},\ }\href {\doibase 10.1103/PhysRevA.55.R1561} {\bibfield  {journal}
  {\bibinfo  {journal} {Phys. Rev. A}\ }\textbf {\bibinfo {volume} {55}},\
  \bibinfo {pages} {R1561} (\bibinfo {year} {1997})}\BibitemShut {NoStop}%
\bibitem [{\citenamefont {Bogdanov}(2009)}]{Bogdanov_JETP2009}%
  \BibitemOpen
  \bibfield  {author} {\bibinfo {author} {\bibfnamefont {Y.~I.}\ \bibnamefont
  {Bogdanov}},\ }\href {\doibase 10.1134/S106377610906003X} {\bibfield
  {journal} {\bibinfo  {journal} {J. Exp. Theor. Phys.}\ }\textbf {\bibinfo
  {volume} {108}},\ \bibinfo {pages} {928} (\bibinfo {year}
  {2009})}\BibitemShut {NoStop}%
\bibitem [{\citenamefont {Bures}(1969)}]{Bures69}%
  \BibitemOpen
  \bibfield  {author} {\bibinfo {author} {\bibfnamefont {D.}~\bibnamefont
  {Bures}},\ }\href@noop {} {\bibfield  {journal} {\bibinfo  {journal} {Trans.
  Amer. Math. Soc.}\ }\textbf {\bibinfo {volume} {135}},\ \bibinfo {pages} {199
  } (\bibinfo {year} {1969})}\BibitemShut {NoStop}%
\bibitem [{\citenamefont {Krauss}(1983)}]{Krauss_book_83}%
  \BibitemOpen
  \bibfield  {author} {\bibinfo {author} {\bibfnamefont {K.}~\bibnamefont
  {Krauss}},\ }in\ \href@noop {} {\emph {\bibinfo {booktitle} {Lecture Notes in
  Physics}}},\ Vol.\ \bibinfo {volume} {190},\ \bibinfo {editor} {edited by\
  \bibinfo {editor} {\bibfnamefont {K.}~\bibnamefont {Kraus}}, \bibinfo
  {editor} {\bibfnamefont {A.}~\bibnamefont {Böhm}}, \bibinfo {editor}
  {\bibfnamefont {J.~D.}\ \bibnamefont {Dollard}}, \ and\ \bibinfo {editor}
  {\bibfnamefont {W.~H.}\ \bibnamefont {Wootters}}}\ (\bibinfo  {publisher}
  {Springer-Verlag},\ \bibinfo {year} {1983})\BibitemShut {NoStop}%
\bibitem [{Note1()}]{Note1}%
  \BibitemOpen
  \bibinfo {note} {Strictly speaking, the Bloch sphere depicts only the
  normalized matrices~$M$ with $\protect \Tr M = 1$, and the space outside the
  sphere corresponds to negative-definite matrices. However, in Fig.~\ref
  {fig:NormalizationViolation}, we treat the definition of the Bloch sphere
  rather freely, meaning that the operator $M$ and the corresponding vector~$s$
  are tied by a relation: $M = \protect \frac {\protect \Tr M +s \sigma }{2}$,
  where $\sigma $ is the vector of three Pauli matrices.}\BibitemShut {Stop}%
\bibitem [{\citenamefont {Struchalin}\ \emph {et~al.}(2018)\citenamefont
  {Struchalin}, \citenamefont {Kovlakov}, \citenamefont {Straupe},\ and\
  \citenamefont {Kulik}}]{Kulik_PRA18}%
  \BibitemOpen
  \bibfield  {author} {\bibinfo {author} {\bibfnamefont {G.~I.}\ \bibnamefont
  {Struchalin}}, \bibinfo {author} {\bibfnamefont {E.~V.}\ \bibnamefont
  {Kovlakov}}, \bibinfo {author} {\bibfnamefont {S.~S.}\ \bibnamefont
  {Straupe}}, \ and\ \bibinfo {author} {\bibfnamefont {S.~P.}\ \bibnamefont
  {Kulik}},\ }\href {\doibase 10.1103/PhysRevA.98.032330} {\bibfield  {journal}
  {\bibinfo  {journal} {Phys. Rev. A}\ }\textbf {\bibinfo {volume} {98}},\
  \bibinfo {pages} {032330} (\bibinfo {year} {2018})}\BibitemShut {NoStop}%
\bibitem [{\citenamefont {Wootters}\ and\ \citenamefont
  {Fields}(1989)}]{Wootters89}%
  \BibitemOpen
  \bibfield  {author} {\bibinfo {author} {\bibfnamefont {W.~K.}\ \bibnamefont
  {Wootters}}\ and\ \bibinfo {author} {\bibfnamefont {B.~D.}\ \bibnamefont
  {Fields}},\ }\href {\doibase 10.1016/0003-4916(89)90322-9} {\bibfield
  {journal} {\bibinfo  {journal} {Annals of Physics}\ }\textbf {\bibinfo
  {volume} {191}},\ \bibinfo {pages} {363 } (\bibinfo {year}
  {1989})}\BibitemShut {NoStop}%
\bibitem [{\citenamefont {\.Zyczkowski}\ \emph {et~al.}(2011)\citenamefont
  {\.Zyczkowski}, \citenamefont {Penson}, \citenamefont {Nechita},\ and\
  \citenamefont {Collins}}]{Zyczkowski_JMP11}%
  \BibitemOpen
  \bibfield  {author} {\bibinfo {author} {\bibfnamefont {K.}~\bibnamefont
  {\.Zyczkowski}}, \bibinfo {author} {\bibfnamefont {K.~A.}\ \bibnamefont
  {Penson}}, \bibinfo {author} {\bibfnamefont {I.}~\bibnamefont {Nechita}}, \
  and\ \bibinfo {author} {\bibfnamefont {B.}~\bibnamefont {Collins}},\ }\href
  {\doibase http://dx.doi.org/10.1063/1.3595693} {\bibfield  {journal}
  {\bibinfo  {journal} {J. Math. Phys.}\ }\textbf {\bibinfo {volume} {52}},\
  \bibinfo {eid} {062201} (\bibinfo {year} {2011})}\BibitemShut {NoStop}%
\bibitem [{\citenamefont {Bogdanov}\ \emph {et~al.}(2011)\citenamefont
  {Bogdanov}, \citenamefont {Brida}, \citenamefont {Bukeev}, \citenamefont
  {Genovese}, \citenamefont {Kravtsov}, \citenamefont {Kulik}, \citenamefont
  {Moreva}, \citenamefont {Soloviev},\ and\ \citenamefont
  {Shurupov}}]{Bogdanov_PRA2011}%
  \BibitemOpen
  \bibfield  {author} {\bibinfo {author} {\bibfnamefont {Y.~I.}\ \bibnamefont
  {Bogdanov}}, \bibinfo {author} {\bibfnamefont {G.}~\bibnamefont {Brida}},
  \bibinfo {author} {\bibfnamefont {I.~D.}\ \bibnamefont {Bukeev}}, \bibinfo
  {author} {\bibfnamefont {M.}~\bibnamefont {Genovese}}, \bibinfo {author}
  {\bibfnamefont {K.~S.}\ \bibnamefont {Kravtsov}}, \bibinfo {author}
  {\bibfnamefont {S.~P.}\ \bibnamefont {Kulik}}, \bibinfo {author}
  {\bibfnamefont {E.~V.}\ \bibnamefont {Moreva}}, \bibinfo {author}
  {\bibfnamefont {A.~A.}\ \bibnamefont {Soloviev}}, \ and\ \bibinfo {author}
  {\bibfnamefont {A.~P.}\ \bibnamefont {Shurupov}},\ }\href {\doibase
  10.1103/PhysRevA.84.042108} {\bibfield  {journal} {\bibinfo  {journal} {Phys.
  Rev. A}\ }\textbf {\bibinfo {volume} {84}},\ \bibinfo {pages} {042108}
  (\bibinfo {year} {2011})}\BibitemShut {NoStop}%
\bibitem [{Note2()}]{Note2}%
  \BibitemOpen
  \bibinfo {note} {An ability to simultaneously measure two orthogonal
  projectors was also taken into account in the simulations.}\BibitemShut
  {Stop}%
\bibitem [{\citenamefont {Ling}\ \emph {et~al.}(2006)\citenamefont {Ling},
  \citenamefont {Soh}, \citenamefont {Lamas-Linares},\ and\ \citenamefont
  {Kurtsiefer}}]{Ling06}%
  \BibitemOpen
  \bibfield  {author} {\bibinfo {author} {\bibfnamefont {A.}~\bibnamefont
  {Ling}}, \bibinfo {author} {\bibfnamefont {K.~P.}\ \bibnamefont {Soh}},
  \bibinfo {author} {\bibfnamefont {A.}~\bibnamefont {Lamas-Linares}}, \ and\
  \bibinfo {author} {\bibfnamefont {C.}~\bibnamefont {Kurtsiefer}},\ }\href
  {\doibase 10.1103/PhysRevA.74.022309} {\bibfield  {journal} {\bibinfo
  {journal} {Phys. Rev. A}\ }\textbf {\bibinfo {volume} {74}},\ \bibinfo
  {pages} {022309} (\bibinfo {year} {2006})}\BibitemShut {NoStop}%
\end{thebibliography}%

\end{document}